\documentclass[12pt,preprint]{emulateapj}
\usepackage{graphicx}

\shorttitle{A Potential Super-Venus}
\shortauthors{Stephen R. Kane et al.}
\slugcomment{Submitted for publication in the Astrophysical Journal
  Letters}

\begin{document}

\title{A Potential Super-Venus in the Kepler-69 System}
\author{
  Stephen R. Kane\altaffilmark{1},
  Thomas Barclay\altaffilmark{2},
  Dawn M. Gelino\altaffilmark{1}
}
\email{skane@ipac.caltech.edu}
\altaffiltext{1}{NASA Exoplanet Science Institute, Caltech, MS 100-22,
  770 South Wilson Avenue, Pasadena, CA 91125, USA}
\altaffiltext{2}{NASA Ames Research Center, M/S 244-30, Moffett Field,
  CA 94035, USA}


\begin{abstract}

Transiting planets have greatly expanded and diversified the exoplanet
field. These planets provide greater access to characterization of
exoplanet atmospheres and structure. The Kepler mission has been
particularly successful in expanding the exoplanet inventory, even to
planets smaller than the Earth. The orbital period sensitivity of the
Kepler data is now extending into the Habitable Zones of their host
stars, and several planets larger than the Earth have been found to
lie therein. Here we examine one such proposed planet, Kepler-69c. We
provide new orbital parameters for this planet and an in-depth
analysis of the Habitable Zone. We find that, even under optimistic
conditions, this $1.7$~R$_\oplus$ planet is unlikely to be within the
Habitable Zone of Kepler-69. Furthermore, the planet receives an
incident flux of 1.91 times the solar constant, which is similar to
that received by Venus. We thus suggest that this planet is likely a
super-Venus rather than a super-Earth in terms of atmospheric
properties and habitability, and we propose follow-up observations to
disentangle the ambiguity.

\end{abstract}



\section{Introduction}
\label{intro}

The discovery of exoplanets has spawned a rapidly growing field of
astronomy, both in number and diversity. A major contributior to this
growth is the success of the transit technique in detecting
exoplanets. This has not only produced a high yield of detections, but
has enabled many follow-up and characterization studies of these
planets due to the additional information the technique provides, in
particular the radius of the planet. The Kepler mission has greatly
increased the number of confirmed transiting exoplanets, with several
thousand candidate planets awaiting confirmation
\citep{bat13,bor11a,bor11b}. The primary mission goal of Kepler is to
detect Earth-size planets within the Habitable Zone (HZ) of their
parent stars \citep{kop13a}, and thus determine their frequency, which
is often denoted by $\eta_\oplus$. The frequency of terrestrial
planets has already been estimated by several authors
\citep{how12,pet13} and these occurrence rates further applied to the
HZ of FGK stars \citep{tra12} and M stars \citep{dre13,kop13b}.
Additional data will likely need to be acquired before $\eta_\oplus$
is a robustly determined quantity. Even so, Kepler data has resulted
in several promising HZ planets with radii slightly larger than the
Earth \citep{bor12,bor13}.

Along the pathway to determining $\eta_\oplus$, Kepler will first
provide insights into the frequency of Mercury and Venus analogs, both
in terms of planet size and incident flux, which we may refer to as
$\eta_{\mathrm{Mercury}}$ and $\eta_{\mathrm{Venus}}$
respectively. The fact that Kepler photometry is of sufficient
precision to detect Mercury-size planets was recently demonstrated by
\citet{bar13a} who discovered a sub-Mercury size planet in the
Kepler-37 system. Since Venus and Earth are approximately the same
size, the Venus analogs will emerge earlier than the Earth analogs due
to the increased geometric transit probability \citep{kan08}. Although
it is not possible to directly distinguish between the Venus and Earth
analogs in terms of their surface conditions, we can determine whether
the planets lie within the HZ of their stars as well as estimate
equilibrium temperatures.

The recent discovery of two planets in the Kepler-69 system by
\citet{bar13b} presents another possible case for a HZ planet, that of
Kepler-69c. The estimated radius of the planet is $\sim
1.7$~R$_\oplus$ which places it in the category of a ``super-Earth''
type of object. The determination of the HZ was conducted based upon
assumptions regarding planetary albedos and equilibrium
temperatures. However, the planet resides on the hot edge of what one
would consider a habitable region.

Here we perform a re-analysis of this system to determine the HZ
classification of the planets. In Section \ref{orbit}, we provide a
self-consistent Maximum Likelihood Estimate (MLE) Keplerian solution
for both planets and discuss their values in relation to the Markov
Chain Monte Carlo (MCMC) estimations provided by \citet{bar13b}. In
Section \ref{hz} we calculate HZ regions for the star under various
assumptions and show that Kepler-69c is unlikely to be a HZ planet. In
Section \ref{imp} we describe the implications of this result for the
planetary evolution and atmosphere and show that this planet is a
strong candidate to be considered a super-Venus. We also discuss
potential signatures of such a planet which may be used to test the
hypothesis.


\section{The Kepler-69 Orbital Solution}
\label{orbit}

In order to provide an accurate HZ analysis, we first require a
complete and self-consistent model of the star and the planetary
orbits. For the star, we adopt the stellar properties derived by
\citet{bar13b}. These were extracted from Keck I HIRES spectra through
both an SME analysis \citep{val05} and a comparison with a library of
stellar spectra. The relevant stellar parameters are shown in the left
side of Table \ref{paramtab}.

\noindent
\begin{deluxetable*}{lc|lcccc}
  \tablecolumns{7}
  \tablewidth{0pc}
  \tablecaption{\label{paramtab} Kepler-69 System Paramaters}
  \tablehead{
    \multicolumn{2}{c}{\bf{STAR}} &
    \multicolumn{5}{c}{\bf{PLANETS}} \\
    \colhead{} &
    \colhead{} &
    \colhead{} &
    \multicolumn{2}{c}{\em{MCMC Analysis}} &
    \multicolumn{2}{c}{\em{Maximum Likelihood (MLE)}} \\
    \colhead{Parameter} &
    \colhead{Value} &
    \colhead{Parameter} &
    \colhead{Kepler-69b} &
    \colhead{Kepler-69c} &
    \colhead{Kepler-69b} &
    \colhead{Kepler-69c}
  }
  \startdata
$T_{\mathrm{eff}}$ (K) & $5638 \pm 168$            & Period (days)   & $13.722341^{+0.000035}_{-0.000036}$ & $242.4613^{+0.0059}_{-0.0064}$ & $13.7223566^{+0.0000534}_{-0.0000766}$ & $242.467216^{+0.004784}_{-0.004784}$ \\
$\log g$               & $4.40 \pm 0.15$           & $R_p/R_*$       & $0.02207^{+0.00023}_{-0.00018}$     & $0.0168^{+0.00052}_{-0.00052}$ & $0.02196^{+0.00104}_{-0.00016}$        & $0.01638^{+0.00192}_{-0.00038}$      \\
Mass ($M_\odot$)       & $0.810^{+0.090}_{-0.081}$ & $e \cos \omega$ & $0.02^{+0.19}_{-0.14}$              & $-0.01^{+0.14}_{-0.16}$        & $0.000076^{+0.50}_{-0.50}$             & $0.0060^{+0.60}_{-0.58}$ \\
Radius ($R_\odot$)     & $0.93^{+0.18}_{-0.12}$    & $e \sin \omega$ & $-0.07^{+0.09}_{-0.14}$             & $-0.02^{+0.08}_{-0.15}$        & $0.000039^{+0.19}_{-0.41}$             & $0.0071^{0.21}_{-0.51}$ \\
Luminosity ($L_\odot$) & $0.80^{+0.37}_{-0.22}$    & $e$             & $0.16^{+0.17}_{-0.001}$             & $0.14^{+0.18}_{-0.10}$         & 0.0 (fixed)                            & 0.0 (fixed) \\
  \enddata
\end{deluxetable*}

\begin{figure*}
  \begin{center}
    \begin{tabular}{cc}
      \includegraphics[width=6.5cm]{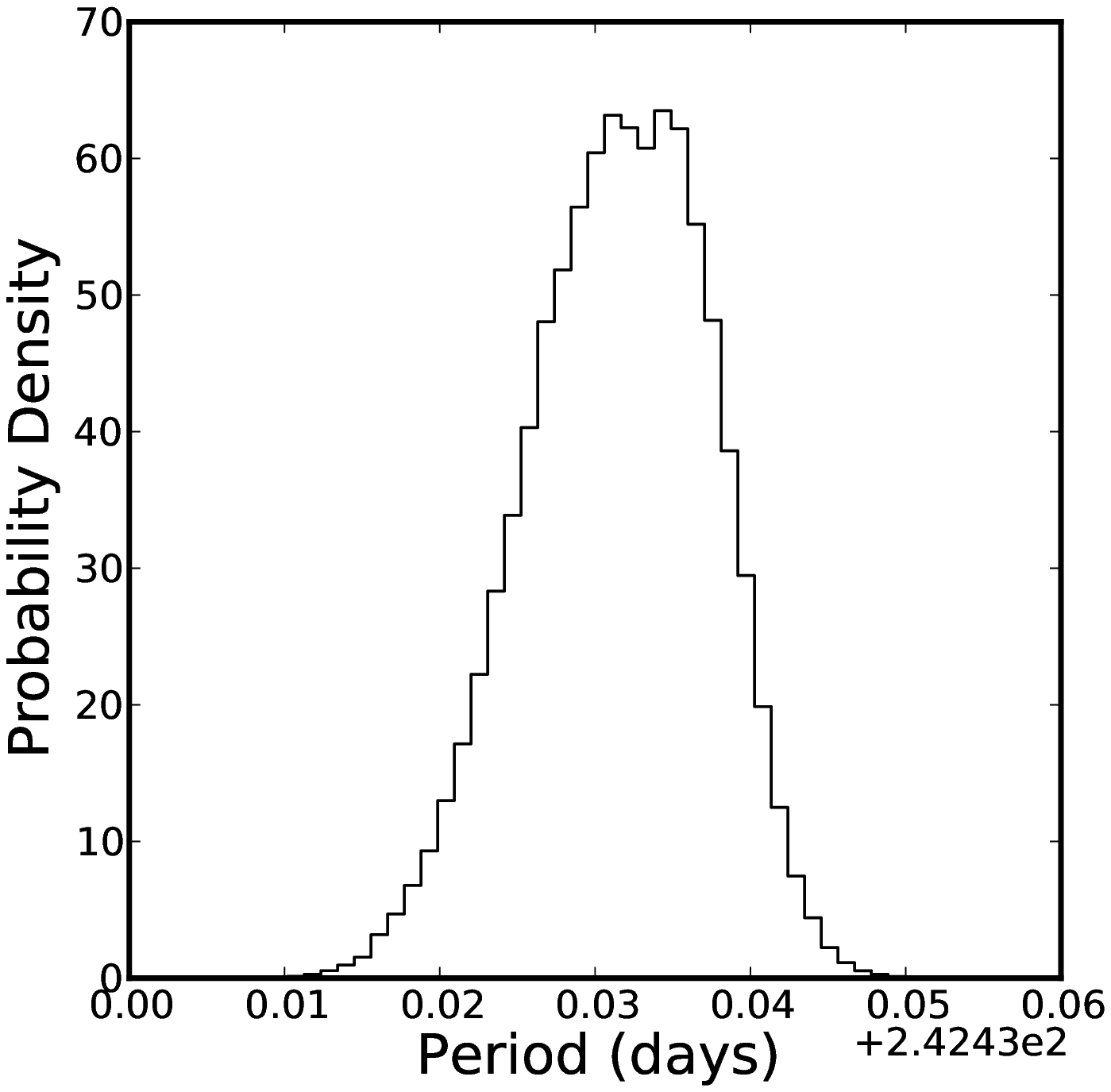} &
      \includegraphics[width=6.5cm]{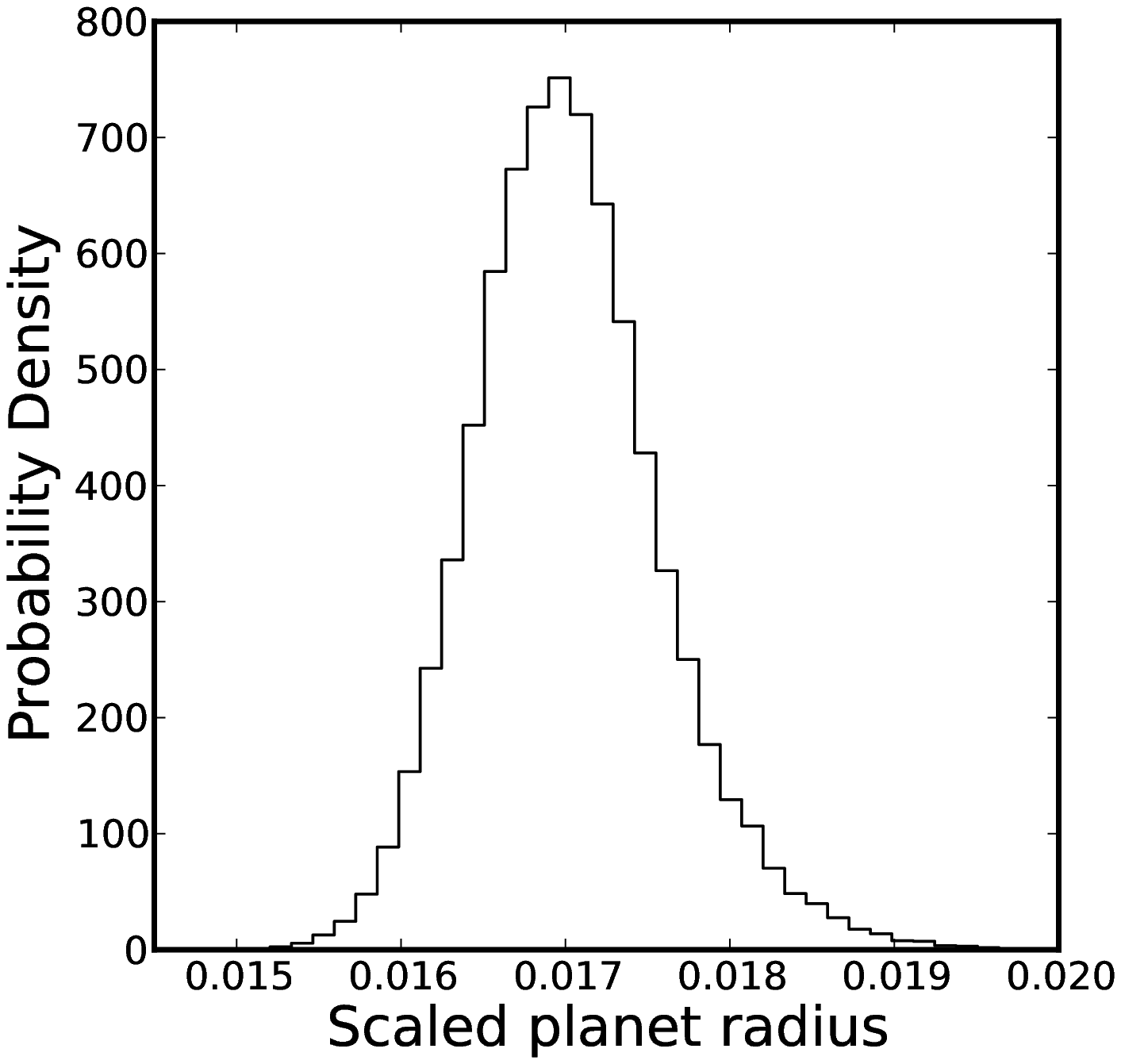} \\
      \includegraphics[width=6.5cm]{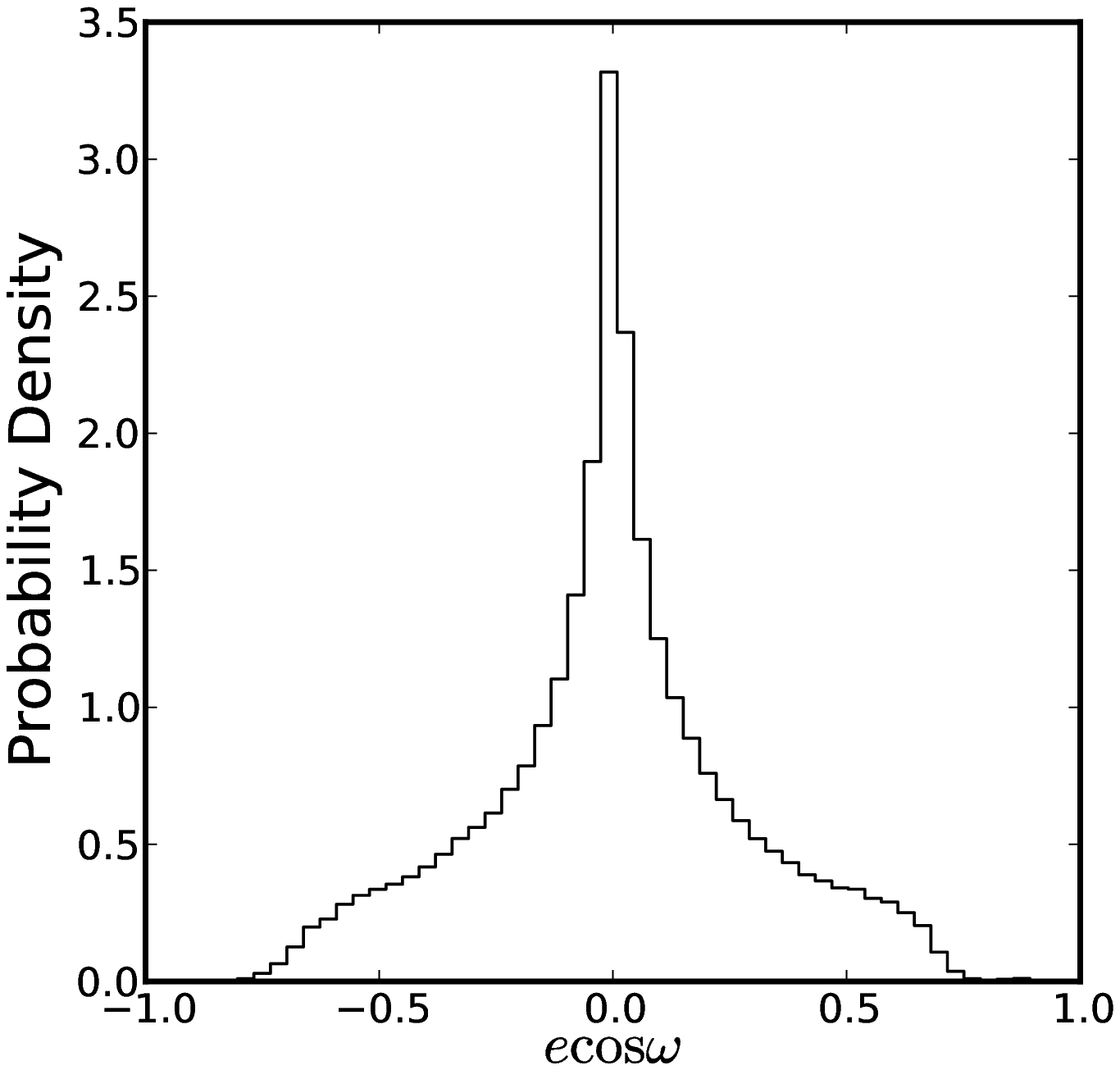} &
      \includegraphics[width=6.5cm]{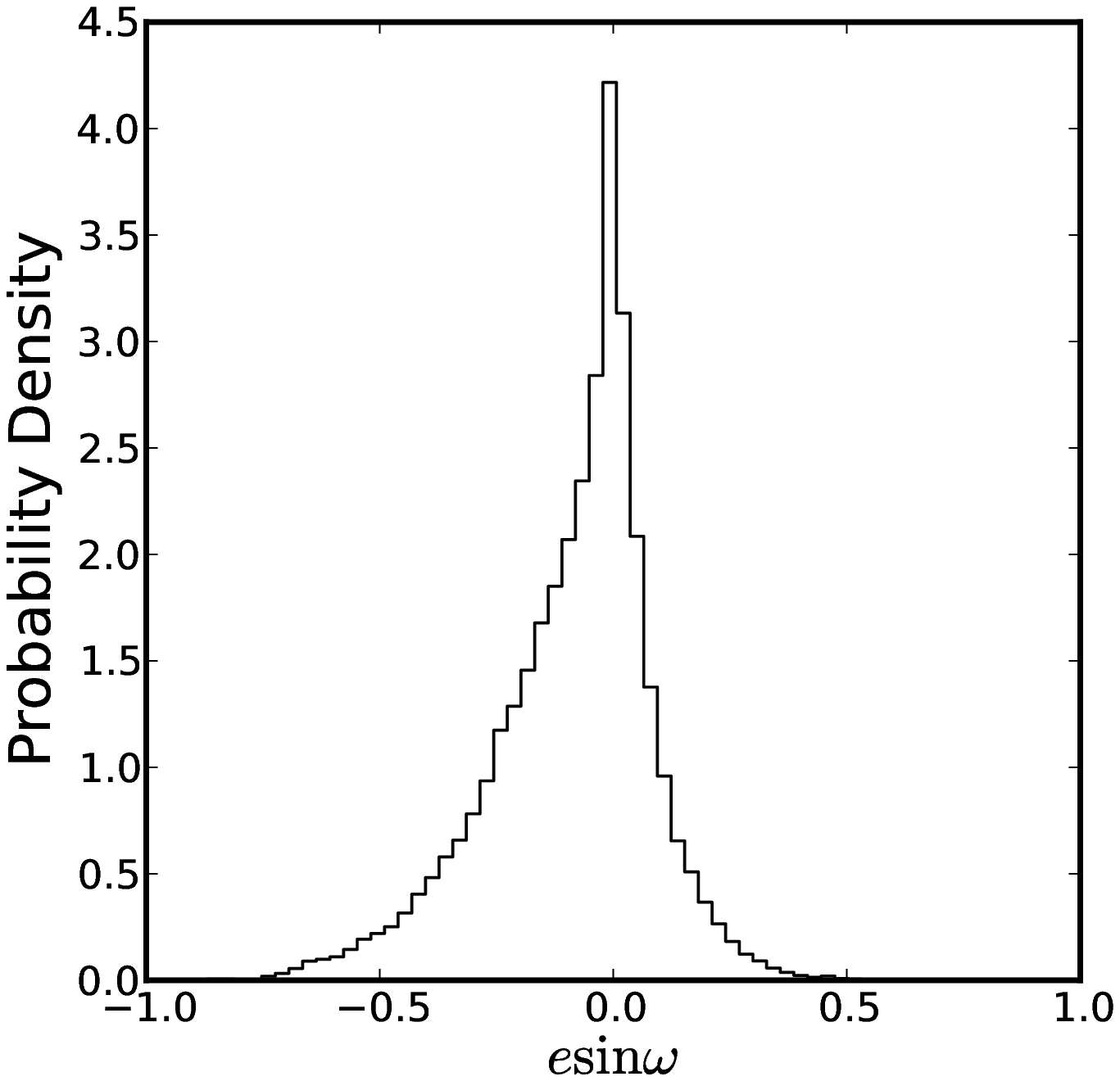}
    \end{tabular}
  \end{center}
  \caption{The probability density distributions for the MLE
    Kepler-69c parameters shown in Table \ref{paramtab}, including the
    orbital period (top-left), $R_p/R_*$ (top-right), $e \cos \omega$
    (bottom-left), and $e \cos \omega$ (bottom-right).}
  \label{orbitarray}
\end{figure*}

\begin{figure*}
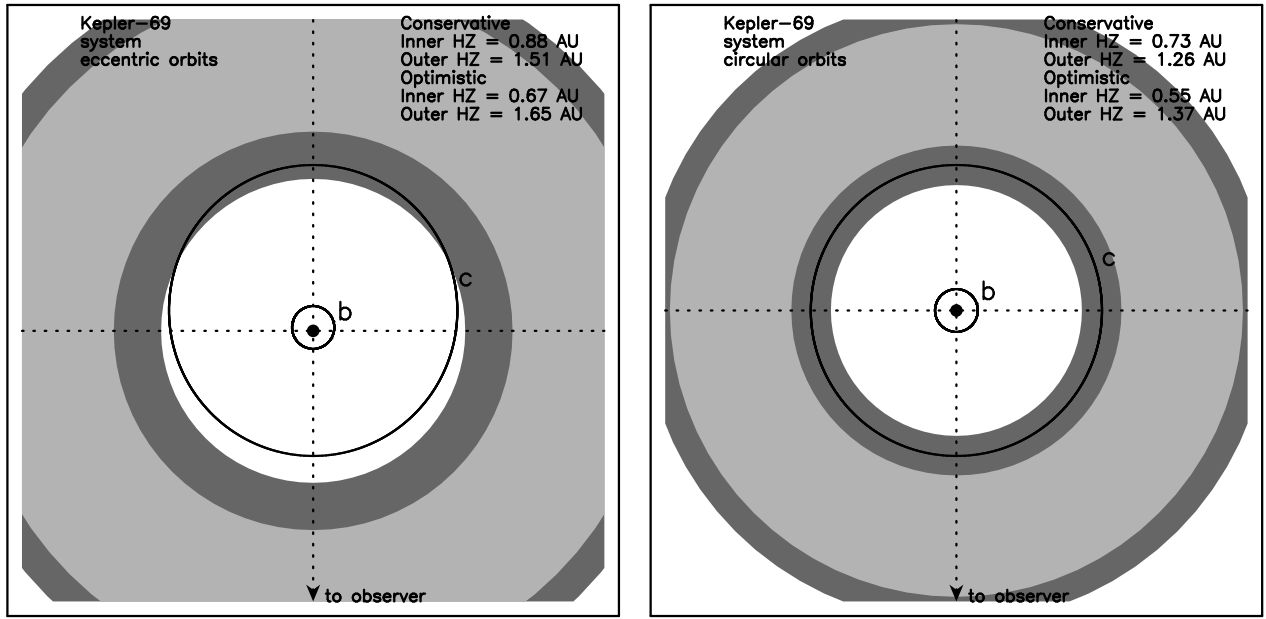

  \begin{center}
    \begin{tabular}{cc}
      \includegraphics[angle=270,width=8.2cm]{f02a.ps} &
      \includegraphics[angle=270,width=8.2cm]{f02b.ps}
    \end{tabular}
  \end{center}
  \caption{The calculated extent of the conservative (light-gray) and
    optimistic (dark-gray) HZ for the Kepler-69 system along with the
    Keplerian orbits of the planets. Left Panel: HZ using the stellar
    parameters and \citet{bar13b} orbital solutions shown in Table
    \ref{paramtab}. Right panel: HZ using Table \ref{paramtab} stellar
    parameters reduced by $1\sigma$ and the circular orbital solution
    presented in this paper.}
  \label{hzfig}
\end{figure*}

The orbital parameters for the planets provided by \citet{bar13b} were
estimated using a MCMC algorithm \citep{for13} with an implicit prior
on eccentricity $e$ imposed by the fit parameters of $e \cos \omega$
and $e \sin \omega$, where $\omega$ is the argument of periastron
\citep{eas13}. This is corrected for using a $1/e$ prior on $e$ and
the final parameters reported are the median values of the posterior
distributions. A prior on the mean stellar density is determined from
stellar evolution models using the spectroscopic parameters. The
parameters resulting from the MCMC analysis are shown in Table
\ref{paramtab}. A problem with this approach of reporting values is
that the final Keplerian solution is not necessarily self-consistent
with respect to the orientation and ellipticity of the orbit. For
example, solving $e \cos \omega$ and $e \sin \omega$ for $e$ using the
MCMC median values yields $e = 0.07$ and $e = 0.02$ for planets b and
c respectively. The Keplerian components of the orbital solution are
important for the HZ calculations described in Section \ref{hz}
because the equilibrium temperature of both planets may vary
substantially during the course of a single orbit due to the
eccentricity, depending on atmospheric circularization acting as a
compensator. Thus a self-consistent solution is required.

Here we report the MLE values from the MCMC analysis. Instead of
reporting the median value for each parameter independently, we report
the single MLE which the MCMC converges upon. This represents both a
single element in the MCMC chain and a self-consistent Keplerian
solution. This solution is shown in Table \ref{paramtab} alongside the
MCMC solution from \citet{bar13b}. The probability density
distribution for each of the parameters from this MLE solution are
shown in Figure \ref{orbitarray}. We also include the uncertainties
for each of the parameters which for a given MCMC chain is the
credible interval which represents 95\% of the parameter
distribution. One thing to note is the relatively large uncertainties
associated with the $e \cos \omega$ and $e \sin \omega$ terms. The
determination of $e$ and $\omega$ are intricately linked to the impact
parameter of the transit crossing and the transit duration
\citep{kan12c}. The transit durations for eccentric and circular
orbits are related by the following equation:
\begin{equation}
  \frac{t_{\mathrm{ecc}}}{t_{\mathrm{circ}}} = \frac{\sqrt{1-e^2}}{1 +
    e \cos(\omega-90\degr)},
    \label{duration}
\end{equation}
The transit duration is also highly dependent upon the stellar radius
which, as shown in Table \ref{paramtab}, is poorly determined for this
star. The degeneracy between impact parameter and transit duration in
determining eccntricity combined with the relatively high uncertainty
in the stellar radius thus results in uncertainty associated with the
values of $e$ and $\omega$. Our presented solution is consistent with
a zero eccentricity. In the following sections we investigate both
circular orbits and the eccentricities found by \citet{bar13b}.


\section{The System Habitable Zone}
\label{hz}

The HZ is usually defined as the region around a star where water can
exist in a liquid state on the surface of a planet with sufficient
atmospheric pressure. One-dimensional climate models were utilized by
\citet{kas93} to quantify the inner and outer boundaries of the HZ for
various types of main sequence stars. These models have been used by
many investigators and have been re-interpreted as analytical
functions of stellar effective temperature and luminosity by several
authors including \citet{sel07}. The work of \citet{kas93} was
recently replaced by the revised models of \citet{kop13a} which
includes an extension of the methodology to later spectral
types. These calculations are available through the Habitable Zone
Gallery \citep{kan12a}, which provides HZ calculations for all known
exoplanetary systems.

The \citet{kas93} and \citet{kop13a} approach considers conditions
whereby the temperature equilibrium would sway to a runaway greenhouse
effect or to a runaway snowball effect based upon the stellar
effective temperature, stellar flux, and water absorption by the
atmosphere. Their boundaries allow for both conservative and
optimistic scenarios depending on how long it is presumed that Venus
and Mars were able to retain liquid water on their surfaces. Here we
use two models, referred to as the ``conservative'' model and the
``optimistic'' model. The conservative model uses the ``Runaway
Greenhouse'' and ``Maximum Greenhouse'' criteria for the inner and
outer HZ boundaries respectively. The optimistic model uses the
``Recent Venus'' and ``Early Mars'' criteria for the inner/outer HZ
boundaries. These criteria are described in detail by \citet{kop13a}.

Figure \ref{hzfig} shows the orbits of the Kepler-69 planets overlaid
on the calculated HZ regions for the star. The light gray represents
the conservative regions and the dark gray regions represent the
extensions to the HZ from the optimistic model. The left panel uses
the stellar parameters and \citet{bar13b} orbital solutions as shown
in Table \ref{paramtab}. The semi-major axes are 0.094~AU for planet b
and 0.64~AU for planet c. For the conservative model, we calculate
inner and outer HZ boundaries of 0.88~AU and 1.51~AU respectively. For
this model, neither planet ever enters the HZ. For the optimistic
model, the inner and outer HZ boundaries are 0.67~AU and 1.65~AU
respectively. In this case, the apastron of the planet c orbit enters
the HZ, assuming the eccentricity of 0.14 shown in Table
\ref{paramtab}. We calculate that 43.9\% of the total orbital period
is spent in the HZ where it is slowly moving through apastron. Planets
in eccentric orbits which spend only part of their orbit in the HZ
have been previously investigated \citep{kan12b,wil02}. These studies
show that habitability is not necessarily ruled out. However, a
grazing encounter with the optimistic inner boundary does not qualify
the planet to be considered a ``HZ planet'' in this model.

We further investigate a scenario which uses our MLE solution shown in
Table \ref{paramtab} for both planets. We assume that the stellar
parameters are $1\sigma$ lower than their measured values, thus we use
an effective temperature and stellar radius of $T_{\mathrm{eff}} =
5470$~K and $R_* = 0.81$~R$_*$ respectively. This results in the HZ
moving inward slightly. For the conservative model, the inner and
outer HZ boundaries are 0.73~AU and 1.26~AU respectively. For the
optimistic model, the inner and outer HZ boundaries are 0.55~AU and
1.37~AU respectively. This scenario is depicted in the right panel of
Figure \ref{hzfig}, and shows that planet c may feasibly spend 100\%
of the orbit within the inner edge of the optimistic boundary. Recall
that this uses the ``Recent Venus'' criteria \citep{kop13a}, which is
empirically determined from an assumption that Venus once had surface
liquid water but has been dry for at least the past 1 billion years
\citep{sol91}. If we adopt the unmodified stellar parameters then
planet c is always interior to even the optimistic HZ.

There are several intrinsic properties of the planet which may move
the effective HZ closer to the star and thus encompass the planetary
orbit. The first of these is the effect of clouds. As pointed out by
\citet{kop13a}, H$_2$O and CO$_2$ cloud effects are not included in
their models since 3D climate models are required. Attempts have been
made to include specific cloud assumptions in HZ calculations, such as
the work of \citet{sel07}. Scaling the results of \citet{sel07} to the
case of Kepler-69c is unlikely to be valid since they use a 1D model
which assumes an Earth-based atmospheric scale height which will
certainly be different for the larger size of Kepler-69c. As shown by
\citet{for97}, neglect of CO$_2$ clouds may result in an
underestimation of the greenhouse effect. Generally speaking, the
effect of H$_2$O clouds is to move the inner HZ boundary closer to the
star but CO$_2$ clouds result in IR back-scattering thus increasing
the surface temperature and moving the HZ outward. Thus the extent to
which the HZ boundary moves is sensitive to the underlying assumptions
regarding the composition and convection of the planetary atmosphere.

A second property of the planet which may influence the extent of the
HZ is the mass. \citet{kop13a} briefly address this aspect for Mars,
Earth, and super-Earth (10~M$_\oplus$) planets. In particular, they
examine the effect of surface gravity which influences the column
depth of the atmosphere and the subsequent strength of the greenhouse
effect. Their results show that that a larger surface gravity will
move the inner edge inward, but a lower surface gravity will move the
outer edge outward. The estimated radius of Kepler-69c is
1.71~R$_\oplus$ \citep{bar13b}. Using the stellar and planetary
properties shown in Table \ref{paramtab} and the mass/radius/flux
relationships of \citet{wei13}, we estimate a Kepler-69c mass of
2.14~M$_\oplus$. This results in a density of $\rho = 2.36$~g/cm$^3$
and a surface gravity of 0.73 times that of the Earth. This relatively
low surface gravity implies that the mass and size of the planet will
have little effect on the HZ and may in fact move the inner HZ
boundary outward.

Finally, we note that \citet{bar13b} address the issue of the HZ by
calculating the equilibrium temperature of the planet based upon
incident stellar flux and assumptions regarding the Bond albedo. As
pointed out by \citet{kop13a}, this approach is problematic since the
albedo is wavelength dependent and may lead to discrepant values for
the HZ boundaries. Our analysis here shows that Kepler-69c is unlikely
to be in the HZ of the host star, although it is a good candidate to
be confirmed as a super-Venus.


\section{Implications for a Possible Super-Venus}
\label{imp}

Considering that Kepler-69c has a similar orbital period to that of
Venus, we consider here the possability that the planet in fact may
bear a Venusian surface environment at an increased scale. A
particularly compelling consideration is that of the incident
flux. The solar flux received by the Earth is approximately
1365~W/m$^2$ and the flux received by Venus is 2611~W/m$^2$, thus
Venus receives 1.91 times more flux than the Earth. Kepler-69c
receives a flux of 2614~W/m$^2$ from the host star which is also a
factor of 1.91 times the flux received at Earth. This similarity in
incident flux received between Venus and Kepler-69c is a primary
motivator in discussing potential other similarities.

\begin{figure*}
  \begin{center}
    \includegraphics[angle=270,width=15.0cm]{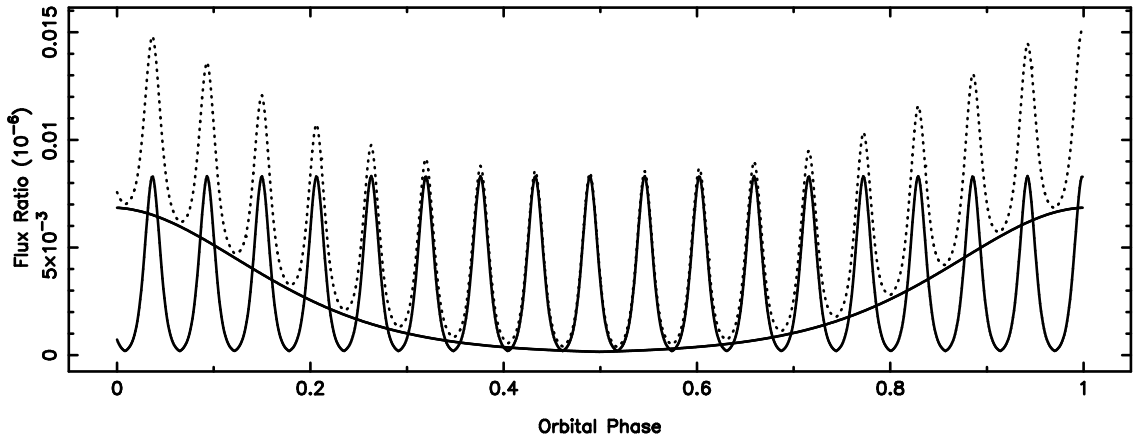}
  \end{center}
  \caption{The model photometric flux variations due to planetary
    phases in the Kepler-69 system for one complete orbital period of
    planet c. The solid lines show the phase variations due to the
    individual planets and the dotted line indicates the combined
    effect.}
  \label{phasefig}
\end{figure*}

The relatively low density of the planet of $\rho = 2.36$~g/cm$^3$ is
similar to the Galilean moons of our Solar System and indicates
composition dominated by silicate and carbonate minerals rather than
metals, consistent with the less than solar metallicity
\citep{bar13b}. Assuming that the planet formed in situ, it will have
been subjected to water delivery during the late heavy bombardment,
although it has been suggested by \citet{mor00} that terrestrial
planets can accrete significant water during formation without the
need for cometary impacts. With such water delivery and a similar
atmospheric evolution to Venus, the availability of silicates and
carbonates in the crust of the planet may well have produced a thick
CO$_2$ atmosphere and subsequent runaway greenhouse warming. In
addition, the non-zero eccentric solution of \citet{bar13b} will help
to trigger further atmospheric changes via tidal heating
\citep{barn13}. Thus a more likely scenario than a habitable
super-Earth with liquid water is an inhospitable planetary surface
with a thick CO$_2$ atmosphere, high temperatures, and high
atmospheric pressure.

While developing techniques to characterize the planets that
contribute to the $\eta_\oplus$ population, such as photometric
variability due to dynamic weather \citep{pal08}, we can similarly
determine how to distinguish these planets from those that comprise
$\eta_{\mathrm{Venus}}$. Although the Venusian surface is hot, the
upper cloud layers are cool and so not necessarily a good indication
of surface conditions. These clouds are also highly reflective, with a
geometric albedo of 0.65 compared with 0.37 for the Earth. This means
that a relatively large amplitude of reflected light (phase)
variations for such planets may be a first indicator of a super-Venus
in exosystems \citep{mal09}. In our own Solar System, Venus is second
only to Jupiter (43\%) in the amplitude of the phase varations
\citep{kan13}. In Figure \ref{phasefig} we show the calculated phase
variations for the Kepler-69 system, including the effects of both
planets for one complete orbital period of planet c. For this model,
we have assumed a high Venusian geometric albedo and a very low albedo
for the inner planet. Since the phase amplitude is proportional to
$R_p^2/a^2$, the inner planet will overwhelmingly dominate the phase
signature. Even so, the phase amplitude of planet c is $\sim 3.5$
times higher than Venus and thus may be a useful diagnostic in similar
systems. Note that the geometric albedo and therefore the phase
amplitude has a strong wavelength dependence.

A more robust method is to measure the major atmospheric components
via transmission spectroscopy \citep{hed13}. The atmospheric signature
of Venus has been extensively studied, and further opportunities via
the recent 2004 and 2012 transits have been exploited to great effect
\citep{hed11}. It is well established that Earth and Venus have many
differences between their transmission spectra, the most prominent
being due to the type of scattering that dominates in the UV and
optical wavelengths \citep{ehr12}. For Earth, Rayleigh scattering is
the dominant effect whereas for Venus it is Mie scattering in the
cloud and haze layers which influences the shape of the spectral
continuum.

Another signature of the Venusian atmosphere is the identification of
the CO$_2$ dominant atmosphere. In emission spectra, CO$_2$ features
emission peaks at 4.3 and 15 $\mu$m due to the atmospheric temperature
inversion \citep{hed13}. Observations of atmopsheric abundances at low
altitudes is difficult for Venus since H$_2$SO$_4$ clouds are
optically thick and obscure the lower atmosphere \citep{sch11,ehr06}.
However, because of the very different temperature structures of the
Earth and Venus atmospheres, the CO$_2$ show significant differences
in transmission spectra. As mentioned earlier, the Venusian upper
atmosphere is efficiently cooled due to the strong greenhouse effect
trapping the IR radiation at the surface. Thus a lack of detactable IR
CO$_2$ emissions in the upper atmosphere may indicate the presence of
a greenhouse closer to the surface which renders the planet
uninhabitable. The detectability of such signatures has been
investigated by \citet{ehr12} and found to be feasible with planned
instrumentation for the James Webb Space Telescope (JWST) and the
European Extremely Large Telescope (E-ELT), although ground-based
detection of CO$_2$ bands is rendered near-impossible due to the
strong presence of the molecule in our own atmosphere. Even so, there
are indeed future instrumentation opportunities that will remove the
ambiguity when attempting to distinguish between super-Earth and
super-Venus atmosphere analogs.


\section{Conclusions}

The extraordinary diversification of the exoplanetary field is
enabling us to finally place our Solar System in context. The
frequency of Jupiter-analogs and Earth-analogs are key components to
understanding the precise nature of that context. However, as the
sensitivity of current detection techniques reaches into the
terrestrial-size regime, we also need to be able to distinguish
between Earth and Venus analogs in order to place our habitability in
context. Thus, the key question of how common Earth-like planets are
relies on also understanding how frequently terrestrial planets
diverge into Venusian-type evolutionary pathways which may place heavy
constraints upon habitability prospects. The Kepler mission is already
revealing many planets which fall into the terrestrial-size planetary
regime and, as we have shown here, there are cases where the
properties of the system straddle current HZ boundary
calculations. The case of the Kepler-69 system is particularly
interesting because the host star is only slightly less massive than
our own sun, thus providing an easier comparison to the HZ and planets
or our Solar System. It is a considerable hinderance to our studies of
larger-than-Earth terrestrial planets that our system conatains no
such objects. However, considering the high occurrance rate of these
types of planets it is likely that these will dominate exoplanet
atmospheric studies well into the next decade. Atmospheric
characterization of large terrestrial planets has a productive future
ahead of it with the anticipated discoveries of the Transiting
Exoplanet Survey Satellite (TESS) and subsequent follow-up with the
JWST. The detection of Venusian-type atmospheres at a variety of
orbital periods and planetary masses will create a picture of where
the divergence between Earth and Venus originates.


\section*{Acknowledgements}

The authors would like to thank David Ciardi, Lauren Weiss, and the
anonymous referee for productive feedback. This work has made use of
the Habitable Zone Gallery at hzgallery.org.



\begin{thebibliography}{}

\bibitem[\protect\citeauthoryear{Barclay et al.}{2013a}]{bar13a}
  Barclay, T., et al. 2013a, Nature, 494, 452
\bibitem[\protect\citeauthoryear{Barclay et al.}{2013b}]{bar13b}
  Barclay, T., et al. 2013b, ApJ, in press (arXiv:1304.4941)
\bibitem[\protect\citeauthoryear{Barnes et al.}{2013b}]{barn13}
  Barnes, R., Mullins, K., Goldblatt, C., Meadows, V.S., Kasting,
  J.F., Heller, R. 2013, AsBio, 13, 225
\bibitem[\protect\citeauthoryear{Batalha et al.}{2013}]{bat13}
  Batalha, N.M., et al., 2013, ApJS, 204, 24
\bibitem[\protect\citeauthoryear{Borucki et al.}{2011a}]{bor11a}
  Borucki, W.J., et al., 2011a, ApJ, 728, 117
\bibitem[\protect\citeauthoryear{Borucki et al.}{2011b}]{bor11b}
  Borucki, W.J., et al., 2011b, ApJ, 736, 19
\bibitem[\protect\citeauthoryear{Borucki et al.}{2012}]{bor12}
  Borucki, W.J., et al., 2012, ApJ, 745, 120
\bibitem[\protect\citeauthoryear{Borucki et al.}{2013}]{bor13}
  Borucki, W.J., et al., 2013, Science, in press (arXiv:1304.7387)
\bibitem[\protect\citeauthoryear{Dressing \&
    Charbonneau}{2013}]{dre13} Dressing, C.D., Charbonneau, D. 2013,
  ApJ, 767, 95
\bibitem[\protect\citeauthoryear{Eastman et al.}{2013}]{eas13}
  Eastman, J., Gaudi, B.S., Agol, E. 2013, PASP, 125, 83
\bibitem[\protect\citeauthoryear{Ehrenreich et al.}{2006}]{ehr06}
  Ehrenreich, D., Tinetti, G., Lecavelier Des Etangs, A.,
  Vidal-Madjar, A., Selsis, F. 2006, A\&A, 448, 379
\bibitem[\protect\citeauthoryear{Ehrenreich et al.}{2012}]{ehr12}
  Ehrenreich, D., et al. 2012, A\&A, 537, L2
\bibitem[\protect\citeauthoryear{Forget \&
    Pierrehumbert}{1997}]{for97} Forget, F., Pierrehumbert, R.T. 1997,
  Science, 278, 1273
\bibitem[\protect\citeauthoryear{Foreman-Mackey et al.}{2013}]{for13}
  Foreman-Mackey, D., Hogg, D.W., Lang, D., Goodman, J. 2013, PASP,
  125, 306
\bibitem[\protect\citeauthoryear{Hedelt et al.}{2011}]{hed11} Hedelt,
  P., et al. 2011, A\&A, 533, A136
\bibitem[\protect\citeauthoryear{Hedelt et al.}{2013}]{hed13} Hedelt,
  P., et al. 2013, A\&A, 553, A9
\bibitem[\protect\citeauthoryear{Howard et al.}{2012}]{how12} Howard,
  A.W., et al. 2012, ApJS, 201, 15
\bibitem[\protect\citeauthoryear{Kane \& von Braun}{2008}]{kan08}
  Kane, S.R., von Braun, K. 2008, ApJ, 689, 492
\bibitem[\protect\citeauthoryear{Kane \& Gelino}{2012a}]{kan12a}
  Kane, S.R., Gelino, D.M. 2012a, PASP, 124, 323
\bibitem[\protect\citeauthoryear{Kane \& Gelino}{2012b}]{kan12b}
  Kane, S.R., Gelino, D.M. 2012b, AsBio, 12, 940
\bibitem[\protect\citeauthoryear{Kane et al.}{2012c}]{kan12c} Kane,
  S.R., Ciardi, D.R., Gelino, D.M., von Braun, K. 2012c, MNRAS, 425,
  757
\bibitem[\protect\citeauthoryear{Kane \& Gelino}{2013}]{kan13} Kane,
  S.R., Gelino, D.M. 2013, ApJ, 762, 129
\bibitem[\protect\citeauthoryear{Kasting et al.}{1993}]{kas93}
  Kasting, J.F., Whitmire, D.P., Reynolds, R.T. 1993, Icarus, 101, 108
\bibitem[\protect\citeauthoryear{Kopparapu et al.}{2013}]{kop13a}
  Kopparapu, R.K., et al. 2013, ApJ, 765, 131
\bibitem[\protect\citeauthoryear{Kopparapu}{2013}]{kop13b} Kopparapu,
  R.K., 2013, ApJ, 767, L8
\bibitem[\protect\citeauthoryear{Mallama}{2009}]{mal09} Mallama,
  A. 2009, Icarus, 204, 11
\bibitem[\protect\citeauthoryear{Morbidelli et al.}{2000}]{mor00}
  Morbidelli, A., Chambers, J., Lunine, J.I., Petit, J.M., Robert,
  F., Valsecchi, G.B., Cyr, K.E. 2000, M\&PS, 35, 1309
\bibitem[\protect\citeauthoryear{Pall\'e et al.}{2008}]{pal08}
  Pall\'e, E., Ford, E.B., Seager, S., Monta\~n\'es-Rodr\'iguez, P.,
  Vazquez, M. 2008, ApJ, 676, 1319
\bibitem[\protect\citeauthoryear{Petigura et al.}{2013}]{pet13}
  Petigura, E.A., Marcy, G.W., Howard, A.W. 2013, ApJ, in press
  (arXiv:1304.0460)
\bibitem[\protect\citeauthoryear{Schaefer \& Fegley}{2011}]{sch11}
  Schaefer, L., Fegley, B. 2011, ApJ, 729, 6
\bibitem[\protect\citeauthoryear{Selsis et al.}{2007}]{sel07} Selsis,
  F., Kasting, J.F., Levrard, B., Paillet, J., Ribas, I., Delfosse,
  X., 2007, A\&A, 476, 1373
\bibitem[\protect\citeauthoryear{Solomon \& Head}{1991}]{sol91}
  Solomon, S.C., Head, J.W. 1991, Science, 252, 252
\bibitem[\protect\citeauthoryear{Traub}{2012}]{tra12} Traub, W. 2012,
  ApJ, 745, 20
\bibitem[\protect\citeauthoryear{Valenti \& Fischer}{2005}]{val05}
  Valenti, J.A., Fischer, D.A. 2005, ApJS, 159, 141
\bibitem[\protect\citeauthoryear{Weiss et al.}{2013}]{wei13} Weiss,
  L.M., et al. 2013, ApJ, 768, 14
\bibitem[\protect\citeauthoryear{Williams \& Pollard}{2002}]{wil02}
  Williams, D.M., Pollard, D. 2002, IJAsB, 1, 61

\end{thebibliography}
\end{document}